\documentclass[12pt]{article}
\usepackage{epsfig}
\usepackage [small]{caption}
\newcommand{\ra}{\rightarrow}
\newcommand{\bq}{\begin{eqnarray}}
\newcommand{\eq}{\end{eqnarray}}
\newcommand{\ov}{\overline}

\begin{document}

\begin{center}{\bf Selected topics in {\boldmath $e^+e^-$} - collisions}
\footnote{\,\,\,
Talk given at the International Workshop $"e^+e^-$ collisions from $\phi$ to $J/\psi"$,\\
March 1, 2006, Novosibirsk, Russia.}\end{center}
\vspace{0.3cm}
\begin{center}{\bf Victor L. Chernyak,}\end{center}
\begin{center} {Budker Institute of Nuclear Physics,
630090 Novosibirsk, Russia,\\ E-mail: v.l.chernyak@inp.nsk.su}
\end{center}

\begin{center} Content \end{center}
1) The leading twist pion wave function\\
2) "Improved" QCD sum rules with non-local condensates\\
3) Pion and kaon form factors and charmonium decays:\\
{\hspace*{3cm}} theory vs experiment\\
4) $\gamma^{*}\gamma\pi^{o}$ - form factor\\
5) The new non-local axial anomaly\\
6) Cross sections $\gamma\gamma\ra \pi^+\pi^-,\, K^+K^-,\,  K_S K_S $\\


{\bf 1. The leading twist pion wave function}\\

The general formula for the leading term of the hadron form factor in QCD was
first obtained in \cite{Ch1} and has the form\,:
\bq
\langle p_2,\,s_2,\,\lambda_2|J_{\lambda}|p_1,\,s_1,\,\lambda_1\rangle =C_{12}
\Bigl (\frac{1}{\sqrt {q^2}}\Bigr )^{|\lambda_1-\lambda_2|+2n_{min}-3}\,,
\eq
where $q=(p_2-p_1),\, n_{\min}$ is the minimal number of elementary constituents in a given
hadron, $n_{min}=2$  for mesons and $n_{min}=3$  for baryons, $s_{1,2}$ and $\lambda_
{1,2}$ are hadron spins and helicities, the current helicity $\lambda=(\lambda_1+\lambda_2)=
0,\,\pm 1$, and the coefficient $C_{12}$ is expressed through the integral over the wave
functions of both hadrons. It is seen that the behavior is independent of hadron spins, but
depends essentially on their helicities.

The largest form factors occur only for $\lambda_1=\lambda_2=0$ mesons and $\lambda_1=\lambda
_2=\pm 1/2$ baryons. For two mesons, for instance:
\bq
F_{1,2}(Q^2)=F_{1,2}^{(lead)}(Q^2)\Biggl
(1+O(\alpha_s)+O(\Lambda^2_{QCD}/Q^2) \Biggr ), \nonumber
\eq
\bq
|Q^2 F^{(lead)}_{1,2}(Q^2)|= \frac{8\pi\,{\ov {\alpha}_s}}{9}\,\Bigl |f_{1}f_{2}\int_0^1
\frac{dx}{x}\,\phi_{1}(x,\,{\ov \mu})\int_0^1 \frac{dy}{y}\,\phi_{2}(y,{\ov \mu})\Bigr |\,,
\eq
where $f_{1,2}$ are meson coupling constants,\, for instance, $f_{\pi}\simeq 130\, {\rm MeV}$,
\, $f_K=160\,{\rm MeV}$ etc.,\, and $\phi_{1,2}(x,{\ov \mu)}$ are their leading
twist wave functions which determine distribution of two meson quarks in fractions $x$ and
$1-x$ of the meson momentum, they are normalized as\,: $\int dx\,\phi_i(x,{\ov \mu})=1.$

The properties of the pion wave function $\phi_{\pi}(x,\mu)$ were investigated in \cite{Ch2}
using QCD sum rules \cite{SVZ}. It was obtained that a few lowest moments of $\phi_{\pi}(x,
\mu\sim 1\,{\rm GeV})$ are significantly larger than for the reference ("asymptotic") wave
function $\phi^{(asy)}(x)=6x(1-x)$, and based on these results the model form was proposed:
$\phi_{\pi}^{(CZ)}(x,\mu_o)=30x(1-x)(2x-1)^2$, at the normalization scale $\mu_o=0.5\,{\rm
GeV}.$ This wave function is much wider than $\phi^{(asy)}(x)$, and this increases greatly
calculated values of various amplitudes with pions.

Recently there appeared first reliable lattice data on the value of $a_2^{\pi}(\mu=1\,{\rm
GeV})$, which is the second Gegenbauer moment of $\phi_{\pi}(x,\mu=1\,{\rm GeV})\,$:
\footnote{ \,\,\, The original lattice results for $a_2(\mu)$ are obtained for $\mu^2=7.1\,
{\rm GeV}^2 $ in \cite{lat1} and $\mu^2=5\,{\rm GeV}^2 $ in \cite{lat2}. They are evolved
to $\mu^2=1\,{\rm GeV}^2 $ at NLO (see e.g. \cite{BB}).

Besides, the results for $\langle \xi^2\rangle_{\pi}$ in \cite{lat2} are  really obtained for
the "pions" with the masses $550\,{\rm MeV}<\mu_{\pi}< 1100\,{\rm MeV}$, and extrapolated then
to the chiral limit $\mu_{\pi}\ra 0$. It looks strange at the first sight that the values of
$\langle \xi^2\rangle_{\pi}$ stay in \cite{lat2} nearly intact in the whole interval $0<\mu_
{\pi}<1.1\,{\rm GeV}$, while one can expect that the wave function of the heavy
"pion" with $\mu_{\pi}=1.1\,{\rm GeV}$ will be noticeably narrower (i.e. has smaller
$\langle \xi^2\rangle $) than those of the real pion.}\,\,:
\bq{\hspace*{-1cm}}
\rm {a_2^{\pi}(\mu=1\,GeV)\,\,}
=\left
\{\begin{array}[c]{l} \simeq 0. \quad \rm{for}\quad
\phi_{\pi}(x)\simeq \phi^{\rm{(asy)}}(x)
\\   \hline  \\
\simeq 0.50 \quad \rm{for}\quad \phi_{\pi}(x)=\phi_{\pi}^{CZ}(x,\mu=1\,GeV)
\quad \cite{Ch2}\\  \hline  \\
( 0.19\pm 0.05 )\,\, {\rm from\,\, "improved"\,\, QCD\,\, sum\,\, rules \,\,with }\\
{\rm non-local\,\, condensates,\,\, A.\, Bakulev\, et.\, al.}\; \cite{Bakulev}
\\ \hline \\
\Bigl ( 0.38 \pm 0.23 ^{+0.11}_{-0.06}\Bigr )\quad \rm {lattice,
\quad L.\, Del\,\,\, Debbio \,\, et. \,\, al.}\quad \cite{lat1}
\\  \hline \\
\Bigl ( 0.364 \pm 0.126 \Bigr )\quad \quad \rm {lattice, \quad M.\,
Gockeler \,\, et. \,\, al.}\quad \cite{lat2} \nonumber
\end{array} \right.
\eq

It is seen that the lattice data disfavor $\phi_{\pi}(x,\mu=1\, {\rm GeV})\simeq \phi_{
\pi}^{(asy)}(x)$ and clearly prefer wider wave function, but it is highly desirable to
increase the accuracy of these lattice calculation (and to decrease $\mu_{\pi}$).\\

{\bf 2. "Improved" QCD sum rules with non-local condensates.}\\

The original (and standard) approach \cite{SVZ} for obtaining QCD sum rules calculates the
correlator of two local currents at small distances expanding it into a power series of
local vacuum condensates of increasing dimension:\,
~\footnote{\,\,
\,In practice, the Fourier transform of eq.(3) is usually calculated, supplied in addition by
a special "Borelization" procedure which suppresses contributions of poorly known higher
dimension terms. This is not of principle importance, but is a matter of technical convenience
and improving the expected accuracy. On account of loop
corrections the coefficients $C_n$ depend logarithmically on the scale $z^2$ (or $q^2$).}
\bq
\langle 0|T J_1(z) J_2(0)|0\rangle =\sum_n  (z^2)^n\,C_n\, \langle 0|O_n(0)|0 \rangle .
\eq
In practical applications this series of power corrections is terminated after first several
terms, so that only a small number of phenomenological parameters $\langle 0|O_n|0\rangle,\,
n<n_o$ determine the behavior of many different correlators.
This standard approach was used in \cite{Ch2} to calculate a few lowest moments of the
pion wave function: $\langle \xi^{2n}\rangle_{\pi}=\int_0^1 dx\,\phi_
{\pi}(x)(2x-1)^{2n}.$ It appeared that the values of these moments are larger significantly
than those for $\phi^{(asy)}(x)$. The most important power corrections in these sum rules
originated from the quark condensate $\sim \langle 0|{\ov q}(0)q(0)|0\rangle ^2 .$

The "improved" approach \cite{non-local}\cite{MR}\cite{Bakulev} proposed not to expand a
few lowest dimension non-local condensates, for instance (the gauge links are implied)
$\Phi(z^2)=\langle 0|{\ov q}(0)\Gamma q(z)|0\rangle $ and $\langle 0|G_{\mu\nu}(0)G_{\mu
\nu}(z)|0\rangle$, into a power series in $z^2$, but to keep them as a whole non-local
objects, while neglecting contributions of all other higher dimension non-local condensates.
This is equivalent to keeping in QCD sum rules a definite subset of higher order power
corrections while
neglecting at the same time all other power corrections which are supposed to be small. {\it
This is the basic assumption underlying this "improvement"}. In other words, it was supposed
that the numerically largest contributions to the coefficients $C_n$ in eq.(3) originate
from expansion of a few lowest dimension non-local condensates, while contributions to $C_n$
from higher dimension non-local condensates are small and can be neglected. Clearly, without
this basic assumption the "improvement" has no much meaning as it is impossible to account
for all multi-local condensates. But really, no one justifications of this basic assumption
has been presented in \cite{non-local}\cite{MR}\cite{Bakulev}.

Moreover, within this approach one has to specify beforehand not a few numbers like $\langle
0|G^2(0)|0\rangle,\,\langle 0|{\ov q}(0)q(0)|0\rangle$, but a number of {\it functions}
describing those non-local condensates which are kept unexpanded. Really, nothing definite is
known about these functions, except (at best) their values and some their first derivatives at
the origin. So, in \cite{non-local}\cite{MR}\cite{Bakulev} definite model forms of these
functions were used, which are arbitrary to a large extent. The uncertainties introduced to
the answer by chosen models are poorly controlable. In principle, with such kind of
"improvements" the whole approach nearly loses its meaning, because to find a few {\it pure
numbers} $\langle \xi^{2n}\rangle_{\pi}$ one has to specify beforehand a number of {\it
poorly known functions}.

As for $\langle \xi^{2n}\rangle_{\pi} $, the main "improvement" of the standard sum rules was
a replacement $\langle 0|{\ov q}(0)q(0)|0\rangle ^2\ra \langle 0|{\ov q}(0)q(x){\ov q}(y)q(z)
|0\rangle$, "factorized {\it via} the vacuum dominance hypothesis to the product of two
simplest $\langle 0|{\ov q}(0)q(z)|0\rangle $ condensates" \cite{non-local}\cite{MR}
\cite{Bakulev}.
Clearly, such a "functional factorization" looks very doubtful in comparison with the standard
"one number factorization" of $\langle 0|{\ov q}(0)q(0){\ov q}(0)q(0)|0\rangle$.
Using this approach, it was obtained in the latest paper \cite{Bakulev}:
$\, a_2(\mu=1\,{\rm GeV})\simeq 0.19,\quad a_4(\mu=1\,{\rm GeV})\simeq -0.13$.~
\footnote{\,\,
For a not very clear reason this differs significantly from the previous results obtained
within the same approach in the second paper in \cite{non-local}\,: $\, a_2(\mu=1\,{\rm GeV})
\simeq 0.35,\quad a_4 (\mu=1\,{\rm GeV})\simeq 0.23$.}
This corresponds to the effectively narrow pion wave function, with $\int^1_0 dx\phi_{\pi}(x,
\mu=1\,{\rm GeV})/x=3(1+a_2+a_4)\simeq 3(1.06)$, i.e. nearly the same value as for the
asymptotic wave function with $a_2=a_4=0$.\\

As for the above described basic assumption of "the improved approach" to the QCD sum rules,
we would like to point out that it can be checked explicitly, using such a correlator for
which the answer is known.

Let us consider the correlator of the axial $A_{\mu}(0)={\ov d}(0)\gamma_{\mu}\gamma_5 u(0)$
and pseudoscalar $P(z)={\ov u}(z)i\gamma_5 d(z)$ currents:
\bq
I_{\mu}(z)=\langle 0|T\,A_{\mu}(0)\,P(z)|0\rangle \equiv \frac{z_\mu}{z^4}\,I(z^2)\,,
\eq
\bq
I(z^2)=\sum_{n=0}^\infty (z^2)^n\,C_n\,Z_n \langle 0| O_n(0)|0\rangle_{\mu_o},\;
C_n=\sum_{i=a,b,c...}C_n^i=\sum_i C_n^{(\rm Born,\,i)}\,\Bigl (1+f_n^i\Bigr ),\nonumber
\eq
where $Z_n=Z_n(\mu^2\sim 1/z^2,\mu_o^2)$ are the renormalization factors of the operators
$O_n$, while $f_n^i=O(\alpha_s)$ is due to loop corrections to the hard kernels.

The exact answer for this correlator is well known in the chiral limit $m_{u,d}=0$, so that
calculating it in various approximations one can compare which one is really better, and this
will be a clear check. So, let us forget for a time that we know the exact answer, and let
us calculate this correlator using the "improved" and standard approaches.

The exact analog of the above described basic assumption predicts here that, at each given $n$,
the largest coefficients in eq.(4) originate from the expansion of  lowest dimension non-local
quark condensate $\Phi(z^2)=\langle 0|{\ov q}(0)q(z)|0\rangle$ shown in fig.1a. Decomposing it
in powers of $z^2$, this results in a tower of power corrections with "the largest
coefficients" $ {C}^{(\rm Born,\,a)}_n$:
\bq
I_o^{(\rm fig.1a)}(z^2)= \sum_{n=0}^\infty (z^2)^n\,{C}^{(\rm Born,\,a)}_n\Bigl (1+f_n^a)
Z_n \langle 0| O_n(0)|0\rangle_{\mu_o} \,,
\eq
where $O_n$ are the corresponding local operators, $O_o\sim {\ov q}q,\,\, O_1\sim {\ov q}
\sigma G q$, etc.

The contribution of the fig.1b to $I(z^2)$ in eq.(4) is originally described by the
three-local higher dimension condensate $\langle 0|{\ov q}(0)G(x)q(z)|0\rangle$. Its expansion
produces finally a similar series in powers of $z^2$ which starts from $\langle O_1\rangle$
and have coefficients $C^{(\rm Born,\,b)}_n(1+f_n^b)$. Besides, the diagrams fig.1c,...
(not shown explicitly in fig.1) with additional gluons emitted from the hard quark
propagator in fig.1 produce similar series, starting from higher dimension condensates $\langle
O_n \rangle , n\geq 2, $ and with the coefficients $C^{(\rm Born,\,c)}_n(1+f_n^c)$, etc.
In the framework of the above basic assumption, there should be a clear numerical hierarchy:
$C^{(\rm Born,\,a)}_n \gg C^{(\rm Born,\,b)}_n \gg C^{(\rm Born,\,i)}_n \gg C^{(\rm Born,\,
i+1)}_n\,...\,$ (all $C_n^{(\rm Born,\,i)}$ are parametrically $O(1)$), so that one can retain
only the largest terms $C^{(\rm Born,\,a)}_n$ and safely neglect all others.

Let us recall now that the exact answer for this correlator is very simple (the spectral
density is saturated by the one pion contribution only), and is exhausted by the first term
with $n=0$ in eq.(4).
\begin{figure}{\vspace*{-2.5cm}}
\centering{\hspace{-0.5cm}}
\includegraphics[width=0.7\textwidth]{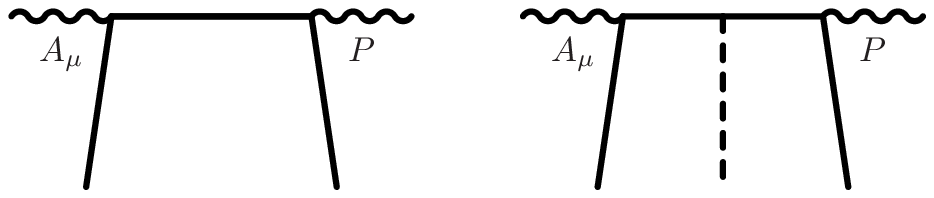}
\caption{}
\end{figure}

In other words, there are no corrections in powers of $(z^2)^{n>0}$ in this correlator at all.
The reason is of course that other contributions with coefficients $C^{(i\neq a)}_n$ in eq.(4)
neglected in the "improved" approach, cancel exactly all (except for the first one) "the most
important" coefficients $C^{(a)}_n$ in eq.(5). For instance, the first power correction $\sim
\langle O_1 \rangle $ from the fig.1b diagram cancels the second term $n=1$ in eq.(5). And
power corrections from other higher dimension multi-local condensates from next diagrams not
shown in fig.1, together with next corrections from the fig.1b diagram, cancel exactly all
next $n\geq 2$ "the most important" terms $C_n^{(a)}$ from the fig.1a diagram in eq.(5).

So, the basic assumption of
the "improved" approach clearly fails: those power corrections which are claimed to be "less
important" in comparison with "the most important corrections", appeared to be not small and
even cancel completely here all "largest corrections".~
\footnote{\,\, When taking the Fourier transform (FT) of eq.(4) the relative values of terms
with different $n$ change, as only singular in $z^2$ terms contribute. So, in this simple
correlator, only two first terms $\langle O_o\rangle$ and $\langle O_1\rangle$ will contribute
to FT of eq.(4) in the lowest order Born approximation: $Z_n=1,\,C_n^i=C_n^{(\rm Born,\,i)}=
{\rm const}_n$. All terms with $n > 1$ contribute to FT only on account of loop corrections.
So, the contributions of $n\geq 2$ terms into FT of eq.(4) will have smaller coefficients
$\sim \alpha_s C_n^{(\rm Born)}$.

In essence, all these technical complications are the extraneous issues for a check of the
basic assumption of "the improved approach", which is essentially the assumed hierarchy
$r_n^{(i)}=(C^{i+1}_n/C^{i}_n) \ll 1$ of coefficients $C_n^i$ {\it at each given $n$} (and
not the relative values of $C_n$ at different $n$)  in the expansion of any
correlator $\langle 0|T J_1(z)J_2(0)|0\rangle$. And this basic assumption can be most easily
and clearly checked just in the co-ordinate representation where the whole tower of terms
survives in the correlator in eq.(4) even in the Born approximation. In any case, $r_n^{(i)}
\simeq C^{({\rm Born},\,i+1)}_n/C^{({\rm Born},\,i)}_n $, independently of whether it is
calculated in the space-time or momentum representations, and so whether $r_n^{(i)}$ is small
or not is independent of the representation and is a real check of the above basic assumption.
Therefore, the statements like: "all terms with $n\geq 2$ enter the FT of eq.(4) with
smaller ($\sim \alpha_s$) coefficients and are invisible in the Born approximation, so that
a check of whether $r_{n\geq 2}^{(i)}$ is small or not can't be performed", look as an
attempt to avoid a real check.

In general, what number of terms in the series in powers of $z^2$ will survive after FT and
in the Born
approximation, depends on the correlator considered. In other correlators more condensates
$\langle O_n\rangle$ will contribute to FT, even in the Born approximation. For instance,
in the sum rules for $\langle \xi^{2k}\rangle_{\pi}$ in \cite{Ch2} all $\langle O_n\rangle$
with ${\rm dim}\, O_n \leq (6+4k)$ will contribute. So, even for $\langle \xi^2\rangle_{\pi}$
there will be a sufficiently large number of various condensates, even in the Born
approximation.

In \cite{Radyu} the appropriate FT of the correlator $\langle 0|T {\ov d}(y)\gamma_{\mu}\gamma
_5 u(-y)\,P(z)|0\rangle$ was considered within the "improved" approach, which resulted in
the sum rules for the pion wave function $\phi_{\pi}(x)$ itself (i.e. for {\it all} moments
$\langle \xi^{2k}\rangle_{\pi}$).
Accounted were only contributions from analogs of the
fig.1a and fig.1b diagrams, while all other contributions were finally neglected, according
to the basic assumption. (Let us note that, even in the Born approximation, all condensates
$\langle O_n\rangle$ with ${\rm dim}\, O_n \leq (5+4k)$ contribute to FT in these sum rules
for the moment $\langle \xi^{2k}\rangle_{\pi}$). Some model form was used for the quark
bi-local condensate $\Phi(v^2)=\langle 0|{\ov q}(0)q(v)|0\rangle$ to describe
the analog of the fig.1a diagram, while the three-local condensate from the analog of the
fig.1b diagram was (in essence, arbitrarily) "simplified" and expressed through the same
$\Phi(v^2)$. The wave function $\phi_{\pi}(x)$ obtained in this way appeared to be
significantly narrower that $\phi^{(\rm asy)}(x)$ (see fig.5 in \cite{Radyu}), with
$a_2^{\pi}\simeq -0.26$ and $\langle \xi^2\rangle_{\pi}\simeq 0.11$ (somewhere at the low
normalization point $\mu\simeq 1\,{\rm GeV}$).
These numbers disagree both with the recent lattice results \cite{lat2}: $a_2^{\pi}(
\mu= 1\,{\rm GeV})=(0.364\pm 0.126)$ and $\langle \xi^2\rangle^{\pi}_{\mu= 1\,{\rm GeV}}
=(0.325\pm 0.043)$, and with previous results of the same author \cite{non-local}. From our
point of view, all this only illustrates that playing with
various "improvements" of sum rules and/or with some model forms for non-local condensates,
one can obtain very different results for $\langle \xi^{2k}\rangle_{\pi}$.
}

On the other hand, calculating the correlator in eq.(4) in the standard approach (which is a
direct QCD calculation accounting for {\it all terms of given dimension}), one finds that the
sum of corrections of given dimension is zero, as it should be.

The conclusion is that the above described "improved approach" to
QCD sum rules can easily give, in general, the misleading results.\\

{\bf 3.\,\, Pion and kaon form factors and charmonium decays: theory vs experiment}\\

The calculated values of $\rm {Br} (\chi_{J}\ra \pi^+\pi^- )$ show high sensitivity to the
precise form of $\phi_{\pi}(x,\mu\sim 1\,{\rm GeV})$ \,\cite{Ch2} (see \cite{CZ} for a
review)\,:
\bq \rm {Br}\Bigl (\chi_o\ra \pi^+\pi^-\Bigr )= \left
\{\begin{array}[c]{l} \simeq 3\cdot 10^{-4}\quad \rm{for}\quad
\phi_{\pi}(x)=\phi^{\rm{(asy)}}(x)
\\   \hline  \\
\simeq 1\cdot 10^{-2}\quad \rm{for}\quad \phi_{\pi}(x)=\phi^{CZ}(x,\mu_o)
\\  \hline  \\
( 0.50\pm 0.06 )\cdot 10^{-2}\quad \rm{experiment}\,\,\, \cite{PDG}
\end{array} \right.
\eq
\bq
\rm {Br}\Bigl (\chi_2\ra \pi^+\pi^-\Bigr )= \left
\{\begin{array}[c]{l} \simeq 1\cdot 10^{-4}\quad \rm{for}\quad
\phi_{\pi}(x)=\phi^{\rm{asymp}}(x)
\\   \hline  \\
\simeq 0.24\cdot 10^{-2}\quad \rm{for}\quad
\phi_{\pi}(x)=\phi^{CZ}(x,\mu_o)
\\  \hline  \\
( 0.18\pm 0.03 )\cdot 10^{-2}\quad \rm{experiment}\,\,\, \cite{PDG}
\end{array}\right.
\eq

It is seen that $\phi_{\pi}^{(CZ)}(x,\mu_o)$ predicts branchings in a reasonable
agreement with the data, while these numbers for $\phi_{\pi}(x)\simeq \phi^{(asy)}(x)$ are
not $\sim 20\%$, but $\sim 20$ times smaller.

The pion form factor, see eq.(2), also shows high sensitivity to the form of $\phi_{\pi}(x,\mu
)$. Below are given some typical numbers for $F_{\pi}^{(lead)}(|Q^2|\simeq Q_o^2=10\,
{\rm GeV}^2)$~\footnote{\,\,
The appropriate choice of ${\ov \alpha_s}$ and $\ov \mu $ in eq.(2) serves to diminish the
role of higher loop corrections to the Born term.
Clearly, the wider is the pion wave function $\phi_{\pi}(x,{\ov \mu})$, the smaller is the
mean virtuality ${\ov \mu}^2\simeq \ov {k^2}$ of the hard gluon in the lowest order Feynman
diagram for $F_{\pi}(Q^2)$, and the larger is $\ov {\alpha_s(k^2)}$.
}
,\, together with the old and recent data.
\bq
\Bigl |\,Q_o^2\,F_{\pi}(Q_o^2)\,\Bigr | = \left \{ \begin{array}[c]{l}
(0.69\pm 0.19)\,,\quad Q^2=9.8\,{\rm GeV}^2\,, \quad \rm {C.J. Bebek\,\, et.\,\,
al.}\: \cite{Bebek} \\   \hline  \\
\simeq 0.13\,\, {\rm GeV}^2\quad \rm{for}\quad
\phi_{\pi}(x)=\phi^{(\rm asy)}(x),\, {\ov \alpha}_s\simeq 0.3 \\  \hline  \\
\simeq 0.5\,\, {\rm GeV}^2\; \rm{for}\quad \phi_{\pi}(x)=\phi^{CZ}(x,\mu_o),\,
{\ov \alpha}_s\simeq 0.4\;\, \cite{Ch2}
\\  \hline  \\
(1.01\pm 0.11\pm  0.07)\,{\rm GeV}^2,\quad \, s=-Q^2=13.48\,{\rm GeV}^2,\\
\hfill CLEO-2005\: \cite{CLEO-ff} \hfill
\end{array}\right.
\eq

\begin{figure}
\centering
\includegraphics[width=0.75\textwidth]{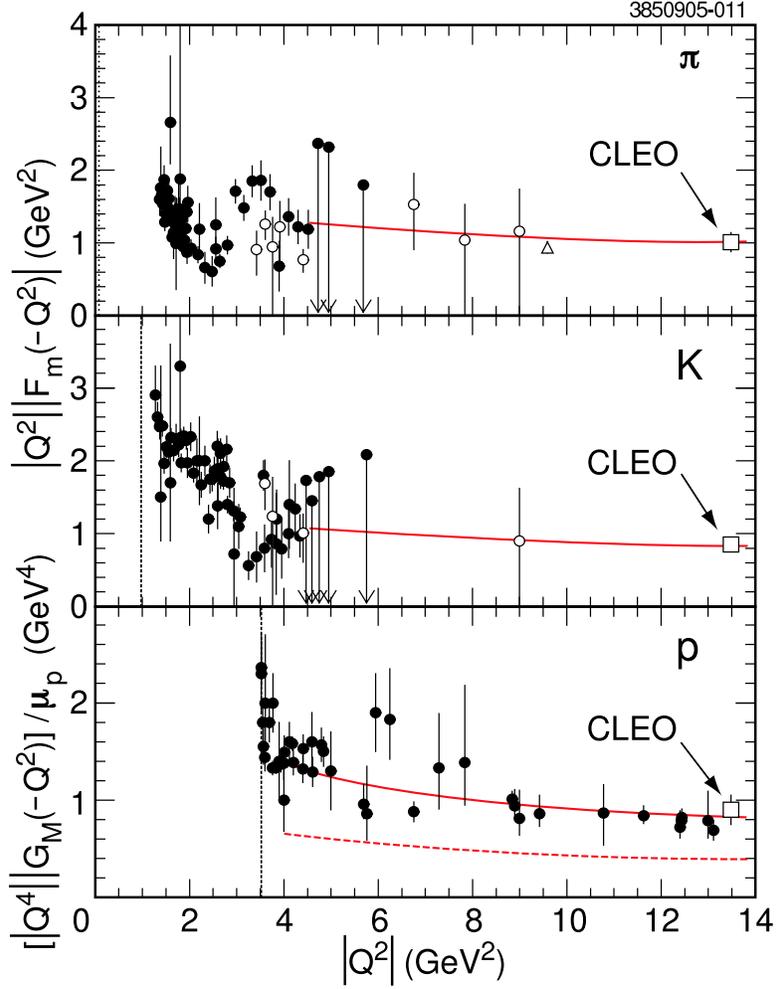}
\caption{
Compilation of the existing experimental data for the
pion, kaon and proton form factors with timelike momentum transfer.
The solid points are from identified $\pi^{\pm}$ and $K^{\pm}$. The
open points are from unidentified $h^{\pm}$, divided into $\pi^{\pm}$
and $K^{\pm}$ according to VDM. The open triangle is from $J/\psi \ra
\pi^+\pi^-$, supposing it proceeds through $J/\psi\ra
\gamma^{*}\ra\pi^+\pi^-$ only. The dashed curve is the fit to the spacelike
form factor of the proton. The new CLEO points \cite{CLEO-ff} are at $s=13.48\,{\rm GeV}^2$.
}
\end{figure}

Estimates show that higher order corrections to the leading term contribution
to  $F_{\pi}(Q^2)$ can constitute up to tens per sent at $Q^2\simeq 10\,{\rm GeV}^2$.
Besides, one naturally expects that the asymptotic behavior is delayed in the time-like region
$s=-Q^2 > 0$, in comparison with the space-like region $s=-Q^2 < 0$, so that, for instance,
$|F_{\pi}(s=10\,{\rm GeV}^2)/F_{\pi}(s=-10\,{\rm GeV}^2)| > 1$, similarly to the nucleon
form factor, see fig.2\,. In any case, this problem with the large value of $|s\,F_{\pi}(s=
13.5\,{\rm GeV}^2)|\simeq 1\,{\rm GeV}^2$ measured by CLEO \cite{CLEO-ff} is much more severe
for $\phi_{\pi}(x)\simeq \phi^{(asy)}(x)$, in which case the leading term contribution has
to be increased $\simeq 7-8 $ times to compare with the data, than for $\phi_{\pi}(x)\simeq
\phi_{\pi}^{(CZ)}(x)$, where the difference is about a factor of two, see eq.(8).

The ratio $F_K(Q^2)/F_{\pi}(Q^2)$ measures the $SU(3)$ symmetry breaking effects. It was
obtained in \cite{Ch3} that the kaon wave function $\phi_K(x)$ is narrower than $\phi_{\pi}(x)
$. This decreases the value of the integral in eq.(2) and compensates for $f_K/f_{\pi}\simeq
1.2$. So, while the naive estimate looks as\,: $F_K(Q^2)/F_{\pi}(Q^2)=(f_K/f_{\pi})^2\simeq
1.5$,\, it was predicted in \cite{CZ}\,: $F_K(Q^2)/F_{\pi}(Q^2)\simeq 0.9$, and this agrees
with the recent CLEO data at $s_o=13.48\,{\rm GeV}^2$\,\cite{CLEO-ff}, see fig.2\,:
\bq
s_o F_{K}(s_o)=(0.84\pm 0.05\pm 0.02){\rm GeV}^2\,,\nonumber
\eq
\bq
s_o F_{\pi}(s_o)=(1.01\pm 0.11\pm 0.07){\rm GeV}^2\,.
\eq

\vspace*{0.5cm}

{\bf 4. The form factor $\gamma^{*}\gamma\pi^{o}.$}\\

This form factor has been measured by CLEO \cite{Savinov} in the range $2\,{\rm GeV}^2 < Q^2
< 8\,{\rm GeV}^2$. The measured value at $Q_o^2=8\,{\rm GeV}^2$ is\,:
\bq
\Phi_{\gamma\pi}(Q_o^2)\equiv Q_o^2 F_{\gamma\pi}(Q_o^2)=(16.7\pm 2.5\pm 0.4)\cdot
10^{-2}\,{\rm GeV}.
\eq

From the theory side, see e.g. \cite{LB},\,\cite{CZ}\,:
\bq
\Phi_{\gamma\pi}(Q^2)= I_{o}\Biggl (1+O(\alpha_s)+O(\Lambda^2_{QCD}/Q^2) \Biggr ),\;
I_{o}=\frac{\sqrt{2}\, f_{\pi}}{3}\int_0^1\frac{dx}{x} \phi_{\pi}(x,\,{\ov \mu}).
\eq

For the asymptotic wave function $\phi_{\pi}^{asy}(x)=6x(1-x):\, I_o^{(asy)}=\sqrt {2}\,f_{
\pi}\simeq 0.184\,{\rm GeV}.$ This is a reference point. For $Q^2=Q_o^2$: a) when using
$\phi_{\pi}^{CZ}(x,{\ov \mu}=0.5\,{\rm GeV})$ in eq.(11), it gives\, $I_o^{(CZ,1)}=(5/3)I_o^
{(asy)}\simeq 0.3\,{\rm GeV}$\,;\, b) when using $\phi_{\pi}^{CZ}(x,{\ov \mu}\simeq 2\,{\rm
GeV}),$ it gives\, $I_o^{(CZ,2)}\simeq (4/3)I_o^{(asy)}\simeq 0.25\,{\rm GeV}.$ It is
seen that, unlike the charmonium decays (see eqs.(6,\,7)) where the decay amplitudes differ by
a factor $\sim 5$ for $\phi_{\pi}^{(asy)}$ and $\phi^{CZ}_{\pi}$, the difference here is only
$\sim 50\% $. So, to infer some conclusions about the form of $\phi_{\pi}(x,\mu)$ from
the CLEO results, it is of crucial importance to estimate reliably
the loop and power corrections in eq.(11).

As for the perturbative loop corrections, the most advanced calculation has been performed by
P. Gosdzinsky and N. Kivel \cite{Kivel} for $\phi_{\pi}(x)=\phi^{(asy)}(x)$. Their result
(calculated in the $b_o\gg 1$ - approximation) looks as: $\Phi_{\gamma\pi}^{(asy)}(Q_o^2)
\simeq I_o^{(asy)}\Bigl ( 1-0.30 \Bigr )\simeq 0.13\,{\rm GeV}$.

As for the $\sim Q^{-2}$  power correction, the value obtained by A. Khodjamirian
\cite{Khodja} for the twist-4 two- and three-particle wave functions contribution is\,:\,
$\Delta \Phi_{\gamma\pi}(Q^2)\simeq {\sqrt 2}\,f_{\pi}(-3\,\delta/Q^2)={\sqrt 2}\,f_{\pi}
(-0.6\,{\rm GeV}^2/Q^2)$. It is seen that it has a typical value $\sim \pm (1\,{\rm GeV}^2/
Q^2)$, expected for a power correction.
So, for $\phi_{\pi}(x)=\phi^{(asy)}(x)$ one has on the whole at $Q_o^2=8\,{\rm GeV}^2$:
\bq{\hspace*{-0.5cm}}
\Phi_{\gamma\pi}^{(asy)}(Q_o^2)\simeq (1-0.30-0.075)\, I_o^{(asy)}=
0.625\cdot 0.184\,{\rm GeV}=0.115\, {\rm GeV} ,
\eq
which looks somewhat small, see eq.(10).

As for $\phi_{CZ}(x,{\ov \mu})$, supposing the approach of \cite{Kivel} will give the relative
values of loop corrections similar to those for $\phi^{(asy)}(x)$, one obtains\,:\, a) when
using $\phi_{\pi}^{CZ}(x,{\ov \mu}=0.5\,{\rm GeV})$ for $I_o$ in eq.(11), this will give\,:\,
$\Phi_{\gamma\pi}^{(CZ,1)}(Q_o^2)\simeq 0.67\cdot 0.3\,{\rm GeV}=0.20\,{\rm GeV} $\,;\, b) when
using $\phi_{\pi}^{CZ,2}(x,{\ov \mu}\simeq2\,{\rm GeV})$, this will give\,:\,$\Phi_{\gamma\pi}
^{(CZ,2)}(Q_o^2)\simeq 0.64\cdot 0.25\,{\rm GeV}=0.16\,{\rm GeV}\,.$ In comparison with the
experimental result in eq.(10), all this is not worse, at least, than for $\phi^{(asy)}(x)$.\\

{\bf 5. The new non-local axial anomaly}.\\

It was claimed by D. Melikhov and B. Stech in \cite{Stech} that the anomaly of the axial
current is not exhausted by the standard local terms, but contains a series of additional
new non-local terms:
\bq
\partial_{\nu}\Bigl ({\bar q}\gamma_{\nu}\gamma_5 q\Bigr )=2m \,\Bigl ({\bar q}\,i\gamma_5\,
q \Bigr ) +Q^2_q\frac{\alpha N_c}{4\pi}\Bigl (F_{\mu\nu}F^{*}_{\mu\nu}\Bigr
)+\frac{\alpha_s}{4\pi}\Bigl (G^a_{\mu\nu}G^{*\,a}_{\mu\nu}\Bigr )+
\nonumber
\eq
\bq +eQ_q\,\kappa\,\frac{1}{\partial^2_{\nu}}\Biggl [ \partial_{\alpha}\Bigl ({\bar q}\gamma_
{\beta}\,q\Bigr )\, F^{*}_{\alpha\beta}\Biggr ]+O(1/\partial^4)\, ,
\eq
where the constant $\kappa\neq 0$.

The explicit example considered in \cite{Stech} from which the above
eq.(13) has been inferred, was the form factor
\bq
T_{\nu}=\langle \rho^{-}(q_2)|A_{\nu}(0)|\gamma(q_1)\rangle\,,
\eq
where: $A_{\nu}={\ov d}\gamma_{\nu}\gamma_5 u$\, is the axial current with the
momentum $p=q_2-q_1$, \, $\gamma(q_1)$ is the on-shell photon and
$\rho^{-}(q_2)$ is the $\rho^{-}$ - meson. This form factor is very
similar to the form factor $\gamma^{*}\gamma\pi^o$ considered in the
previous section.

It has been obtained in \cite{Stech} that (in the chiral limit and
at large $p^2$) the divergence of $A_{\nu}$ is non-zero, see fig.3:
\bq
p_{\nu}\,\langle \rho^{-}(q_2)|A_{\nu}|\gamma(q_1)\rangle \sim
\kappa\,\frac{f_{\rho}M_{\rho}}{p^2}+O \Bigl ( \frac{1}{p^4}\Bigr)\,,
\eq
and the constant $\kappa$ was calculated explicitly through the integral over the $\rho$ -
meson wave functions (see \cite{CZ} for the definitions and asymptotic forms of the $\rho$-
meson wave functions,\, we use here the same
notations and the same asymptotic forms of these wave functions as in \cite{Stech},\,\,
$\Phi(x)=3x(1-x)(2x-1)/2\,,\,\,V_{\perp}(x)=3[1+(2x-1)^2]/4\,,\,\, V_A(x)=6x(1-x)\,\,)$\,:
\bq
\kappa=\int_0^1 dx\, \Bigl [\,\frac{\Phi(x)-(1-x)V_{\perp}(x)}{x}+\frac{(1+x)
V_A(x)}{4\,x^2}\,\Bigr ].
\eq
\begin{figure}
\centering
\includegraphics[width=0.8\textwidth] {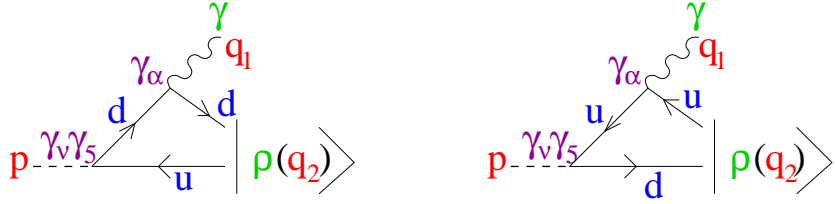}
\caption{The diagrams for the new non-local anomaly}
\end{figure}
Substituting into eq.(16) the above explicit form of the wave functions,
one obtains: $\kappa=-3/2\neq 0$. This is the main result of the paper \cite{Stech}.

This result looks highly surprising because the anomalous contribution originates here directly
from the original light quark operators, see fig.3\,.  One way to make the axial anomaly
"visible", is to use from the beginning the heavy regulator quark fields $\Psi$. Then the
equations
of motion are the standard ones: $\partial_{\nu}\Bigl ({\ov q}\gamma_{\nu}\gamma_5 q +
 {\ov \Psi}\gamma_{\nu}\gamma_5\Psi \Bigr )=2m \,{\ov q}i\gamma_5 q +2M \,{\ov \Psi} i
 \gamma_5 \Psi .$ But the heavy regulator fields $\Psi$ are absent in the external low energy
 states. So, the regulator fields have to be contracted into the loop, from which
only the gauge fields can originate. This gives the standard form of the axial anomaly.
So, the light quarks do not contribute directly to the anomaly (at $m_{u,\,d}=0$
and $q_2^2\neq 0$), and the constant $\kappa$  in eq.(16) should be zero.
On the other hand, the calculation of $\kappa=-3/2$ from
\cite{Stech} looks right, at the first sight. So, what is going wrong ?

Our answer is that the right form of eq.(16) looks really as:
\bq
\kappa=\lim_{\delta \ra 0} \int_0^1 dx\,\Bigl [\, \frac{\Phi(x)-(1-x)V_{\perp}(x)}
{(x+\delta)}+\frac{(1+x)V_A(x)}{4\,(x+\delta)^2}\,\Bigr],
\eq
where $\delta$ is a small power correction $\sim \Lambda^2_{QCD}/p^2$, which is present
in the quark propagator in fig.3 (because quarks inside the $\rho$ - meson are not strictly
on-shell, but have virtualities $\sim \Lambda^2_{QCD}$). Calculating
$\kappa$ from eq.(17) one obtains: $\kappa=0$.

The difference with \cite{Stech} originates clearly from the fact
that in \cite{Stech} the order of $\lim_{\delta\ra 0}$ and $\int^1_0
dx$ was interchanged, but this is not allowed in this case.
\footnote{\hspace*{0.3cm} Another way, one can first integrate by
parts the last term with $V_A(x)$ in eq.(17). The contribution of the
total derivative is zero at $\delta\neq 0$, and this is a crucial
point. After this $\delta$ can be put zero even under the integral,
and one obtains $\kappa=0$ after integration.}

We conclude that there is no any new non-local axial anomaly.\\

{\bf 6. Large angle cross sections $\gamma\gamma\ra \pi^+\pi^-\,,\,
K^+K^-\,,\,\, K_S K_S$.}\\

The leading contributions to the hard kernels for these  amplitudes at large
$s=W^2=(q_1+q_2)^2$ and fixed c.m.s. angle $\theta$ were calculated first in \cite{BL2} for
symmetric meson wave functions, $\phi_M(x)=\phi_M(1-x)$,\, and later in \cite{Maurice}
(BC in what follows) for arbitrary wave functions. Two typical Feynman diagrams are shown
in fig.4\,.
\begin{figure}{\vspace*{-2.5cm}}
\centering{\hspace*{-2cm}}
\includegraphics[width=0.65\textwidth]{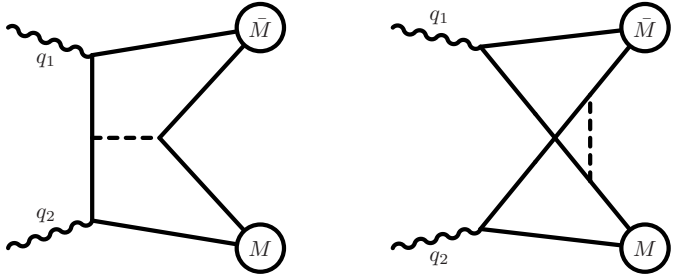}
\caption{Two typical Feynman diagrams for the leading term hard contributions to
$\gamma\gamma\ra {\ov M}M$\,, the broken line is the hard gluon exchange.}
\end{figure}
The main features of these cross sections
are as follows (below we follow mainly the definite predictions from BC in \cite{Maurice})\,.\\
{\bf a)}\,\,  $d\sigma(\pi^+\pi^-)$ can be written as\,:
\bq
\frac{s^3}{16\pi\alpha^2}\,\frac{d\sigma(\gamma\gamma\ra\pi^+\pi^-)}{d |\cos\theta |}
\equiv \frac{|\Phi^{(eff)}_{\pi}(s,\theta)|^2}{\sin ^4 \theta}
=\frac{|s F_{\pi}^{(lead)}(s)|^2}{\sin ^4 \theta}|1- \upsilon(\theta)|^2\,,
\eq
where
$F_{\pi}^{(lead)}(s)$ is the leading term of the pion form factor \cite{Ch1}\,:
\bq
|s F^{(lead)}_{\pi}(s)|= \frac{8\pi\,{\ov {\alpha}_s}}{9}\,\Bigl
|f_{\pi}\int_0^1 \frac{dx}{x}\,\phi_{\pi}(x,\,{\ov \mu})\Bigr |^2\,.
\eq

While the value of $F^{(lead)}_{\pi}(s)$ is very sensitive to the form of $\phi_{\pi}(x)$, the
factor $\upsilon(\theta)$, as emphasized in \cite{BL2}, is nearly independent on the form of
$\phi_{\pi}(x)$, and depends only weakly on $\theta$. Numerically, $\upsilon(\theta)
\simeq 0.12$.

The recent data from the Belle collaboration \cite{Belle1} for $(\pi^+\pi^-)$ and $(K^+K^-)$
agree with $\sim 1/\sin^4\theta $ dependence at $W\geq 3\,{\rm GeV}$, while the angular
distribution is somewhat more steep at lower energies. The energy dependence at $2.4\,{\rm GeV}
 < W < 4.1\,{\rm GeV}$ was fitted in \cite{Belle1} as: $\sigma_o(\pi^+\pi^-)=\int_0^{0.6}dc
 (d\sigma/d|c|)\sim W^{-n}\,,\,\,\,n=(7.9\pm 0.4\pm 1.5)$ for $(\pi^+\pi^-)$, and $n=(7.3\pm
 0.3\pm 1.5)$ for $(K^+K^-)$. The overall value $n\simeq 6 $ is also acceptable however, see
 fig.5\,. As for the absolute normalization, the $(\pi^+\pi^-)$ data are fitted in \cite
{Belle1} to the eq.(18) with\,: $|\Phi_{\pi}^{(eff)}(s,\theta)|=(0.503\pm 0.007\pm 0.035)\,
{\rm GeV}^2$.\,\,\footnote{ \,\,\,Clearly, in addition to the leading terms $A^{(lead)}$ ,
this experimental
value includes also all loop and power corrections $\delta A$ to the $\gamma \gamma\ra \pi^+
\pi^-$ amplitudes $A=A^{(lead)}+\delta A$. These are different of course from corrections
$\delta F_{\pi}$ to the genuine pion form factor $F_{\pi}=F_{\pi}^{(lead)}+\delta F_{\pi}$.
So, the direct connection between the leading terms of $d\sigma(\pi^+\pi^-)$ and
$|F_{\pi}|^2$ in eq.(18) does not hold on account of corrections.}
This value can be compared with \,: $0.88\cdot |s F_{\pi}^{(CZ)}(s)|\simeq 0.4\,{\rm GeV}^2$
for $\phi_{\pi}^{(CZ)}(x,\mu_o)$, and $0.88\cdot |s F_{\pi}^{(asy)}(s)|\simeq 0.12
\,{\rm GeV}^2$ for $\phi^{(asy)}(x).$ It is seen that the wide pion wave function $\phi_{\pi}
^{(CZ)}(x)$ is preferable, while $\phi^{(asy)}(x)$ gives the cross section which is
$\simeq 15$ times smaller than data.  It seems impossible that, at energies $s= 10-15\,
{\rm GeV}^2$, higher loop or power corrections can cure so large difference.
\footnote{\,\, A similar situation occurs in calculations of charmonium decays. ${\rm Br}
(\chi_o\ra \pi^+\pi^-)$ and ${\rm Br} (\chi_2\ra \pi^+\pi^-)$ calculated with $\phi_{\pi}(x)=
\phi^{(asy)}(x)$ are $\simeq 20$ times smaller than the data, while the use of $\phi_{\pi}
(x)=\phi^{(CZ)}(x,\mu_o)$ leads to values in a reasonable agreement with the data, see
eqs.(6,\,7).}
\\

{\bf b)}\,\, The SU(3)-symmetry breaking, $d\sigma(K^+K^-)\neq d\sigma(\pi^+\pi^-),$ originates
not only from different meson couplings, $f_K\neq f_{\pi}$, but also from symmetry breaking
effects in normalized meson wave functions, $\phi_{K}(x)\neq \phi_{\pi}(x)$. These two
effects tend to cancel each other. So, instead of the naive prediction $\simeq (f_K/
f_{\pi})^4\simeq 2.3$ from \cite{BL2}, the prediction of BC for this ratio is close to
unity, and this agrees with the recent data from Belle \cite{Belle1}:
\bq
\frac{\sigma_o (\gamma\gamma\ra K^+K^-)}{\sigma_o (\gamma\gamma\ra
\pi^+\pi^-)}=\left\{\begin{array}[c]{ll} \displaystyle
(f_K/f_{\pi})^4\simeq 2.3 \scriptstyle &
\begin{array}[c]{l}\hspace*{-3cm}\rm{ Brodsky,\, Lepage} \,\, \cite{BL2} \end{array} \\ & \\
 \hline  & \\
\simeq 1.06 & \begin{array}[c]{l} \hspace*{-3.4cm} \rm{Benayoun,\,
Chernyak} \,\,\cite{Maurice}
\end{array} \\ & \\ \hline  & \\
(0.89\pm 0.04\pm 0.15)\quad  {\rm Belle} - 2004 \,\,\, \cite{Belle1}
\end{array}\right. \nonumber
\eq
\begin{figure}{\hspace*{-2cm}}
\includegraphics[width=0.6\textwidth]{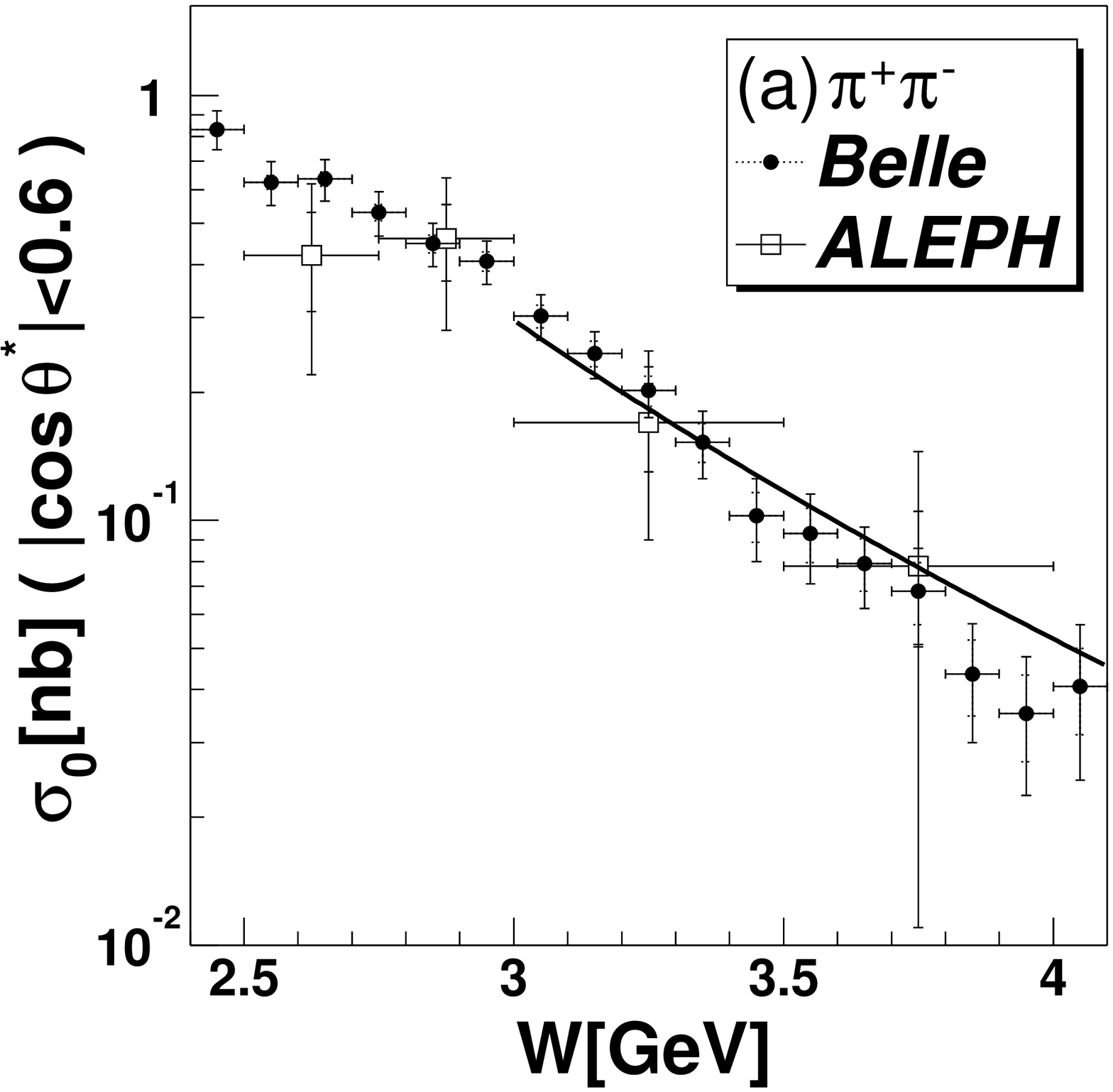}~
\includegraphics[width=0.6\textwidth]{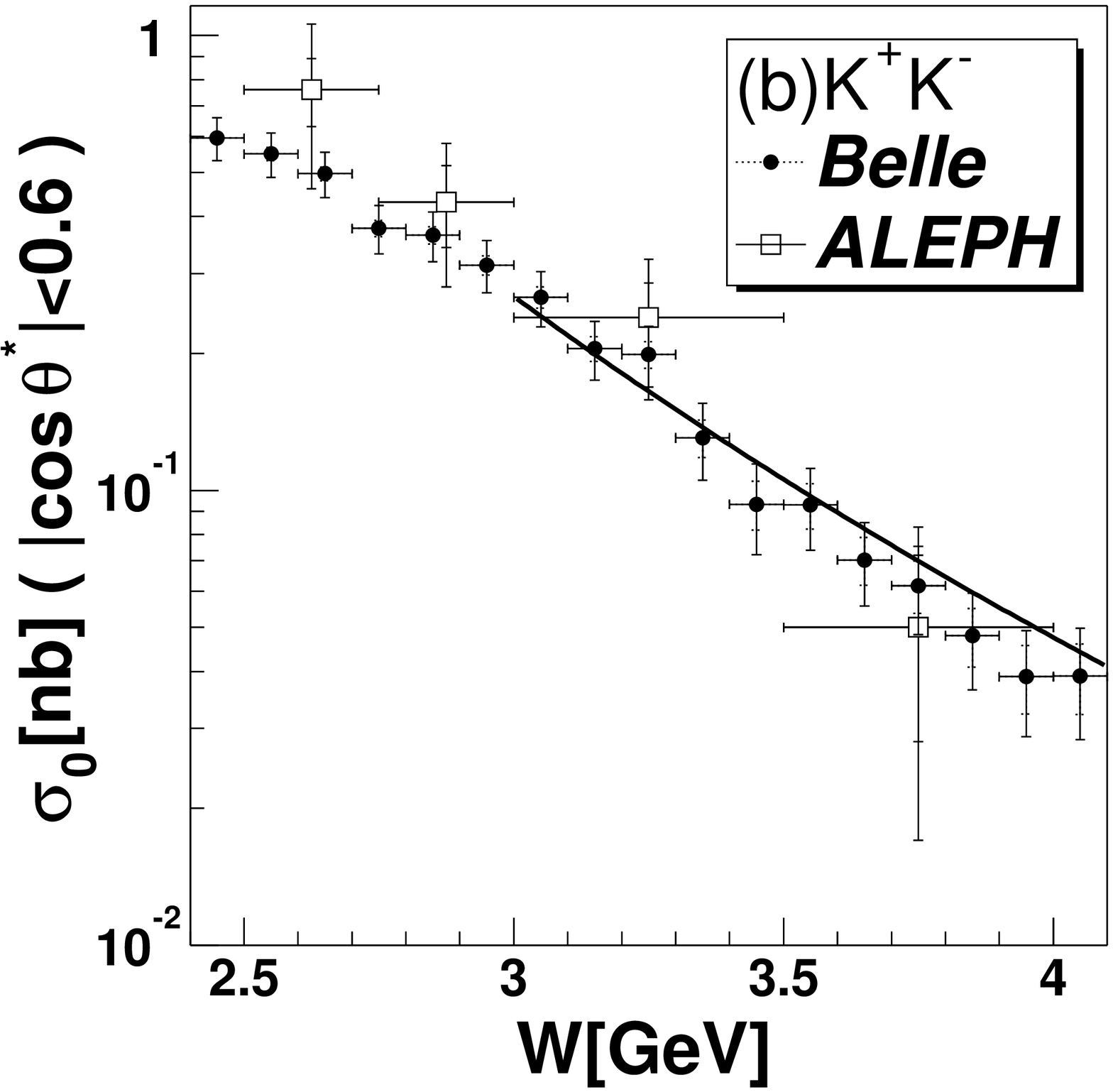}\\
\centering
\includegraphics[width=0.9\textwidth] {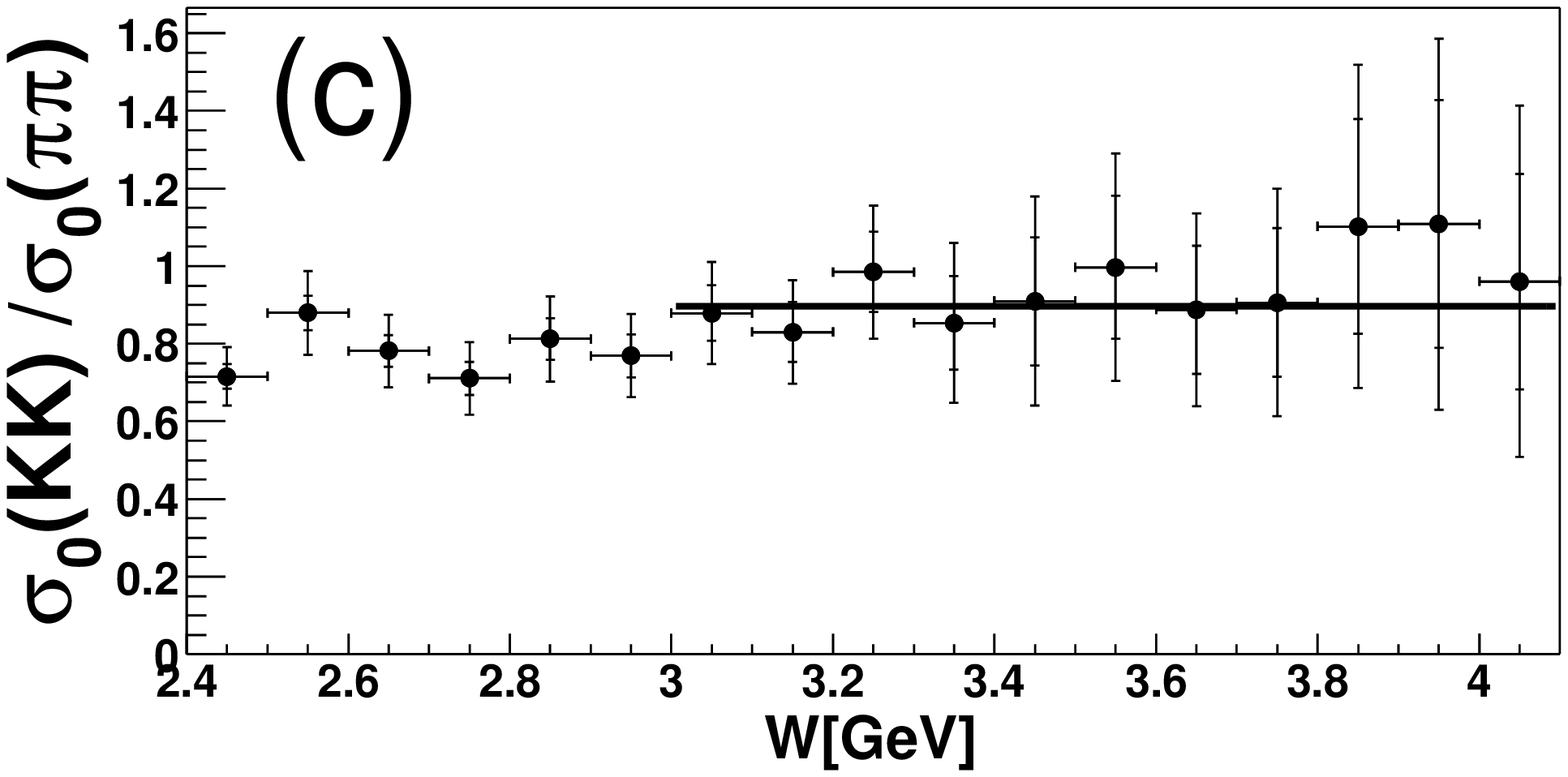}
\caption{ The cross sections $\sigma_o=\int d \cos \theta\,(d\sigma/d\cos\theta)$\,,\,
integrated over the c.m. angular region $|\cos\theta|<0.6\,,\,\,$ for a) $\,\,\gamma\gamma\
\ra \pi^+\pi^-\,\,$, b) $\,\,\gamma\gamma\ra K^+K^-\,\,,$\,
together with a $W^{-6}$ dependence line\,;\,\, c) the cross section
ratio, the solid line is the result of the fit for the data above $3\,{\rm GeV}$\,,\,\,
the errors indicated by short ticks are statistical only.}
\end{figure}~

{\bf c)}\,\, The leading terms in cross sections for neutral particles are much smaller
than for charged ones. For instance, it was obtained by BC that the ratio
$d\sigma^{(lead)}(\pi^o\pi^o)/d\sigma^{(lead)}(\pi^+\pi^-)$ varies from $\simeq 0.07$
at $\cos \theta =0$ to $\simeq 0.04$ at $\cos \theta =0.6$. Besides, it was obtained
therein for the ratio\,: $({\ov K}^o K^o)^{(lead)}/(\pi^o\pi^o)^{(lead)}\simeq 1.3\cdot
(4/25)\simeq 0.21\,.$ So, for instance, one obtains for cross sections $\sigma_o^{(lead)}$
integrated over $0\leq |\cos\theta|\leq 0.6 $ for charged particles and over $0 \leq
\cos\theta \leq 0.6 $ for neutral ones\, :\,\, $\sigma_{o}^{(lead)}(K_S K_S)/\sigma_{o}^
{(lead)}(K^+K^-)\simeq 0.005\,.$

It is seen that the leading contribution to $\sigma_o(K_SK_S)$ is very small. This implies
that, unlike to the case $\sigma_o(K^+K^-)$, it is not yet dominant at present energies $
W^2 < 16\,GeV$. In other words, the amplitude $A(\gamma\gamma\ra K_S K_S)=(a(s,\theta)+b(s,
\theta))$ is dominated by the non-leading term $b(s,\theta)\sim g(\theta)/s^2$, while the
formally leading term $a(s,\theta)\sim C_o f_{BC}(\theta)/s$ has
so small coefficient $C_o$ that $|b(s,\theta)| > |a(s,\theta)|$ at, say, $ W^2 < 12\,GeV^2$.
So, it has no meaning to compare the leading term prediction of BC (i.e. $d\sigma(K_S K_S)/d
\cos\theta\sim |a(s,\theta)|^2/s \sim |f_{BC}(\theta)|^2/W^{6} $ at $s\ra \infty$) for the
energy and angular dependence of $d\sigma(K_S K_S)$ with the recent data from  Belle
\cite{Belle2}. Really, the only QCD prediction for $6\,GeV^2 < W^2 < 12\,GeV^2$ is the energy
dependence: $d\sigma(K_S K_S)/d\cos\theta \sim |b(s,\theta)|^2/s\sim |g(\theta)|^2/W^{10}$,
while the angular dependence $|g(\theta)|^2$ and the absolute normalization are unknown.
This energy dependence agrees with \cite{Belle2}, see fig.7\,.\\

{\bf The hand-bag model} \cite{DKV} (DKV in what follows) is a part of a general ideology
which claims that present day energies are insufficient for the leading terms QCD to be
the main ones. Instead,  the soft nonperturbative contributions are supposed to dominate
the amplitudes. The handbag model represents applications of this ideology
to description of $d\sigma(\gamma\gamma\ra {\ov M}M)$. It assumes that the above described
hard contributions really dominate at very high energies only, while
the main contributions at present energies originate from the fig.6a diagram.
\begin{figure}{\vspace*{-2.5cm}}
\centering{\hspace{-2cm}}
\includegraphics[width=0.9\textwidth]{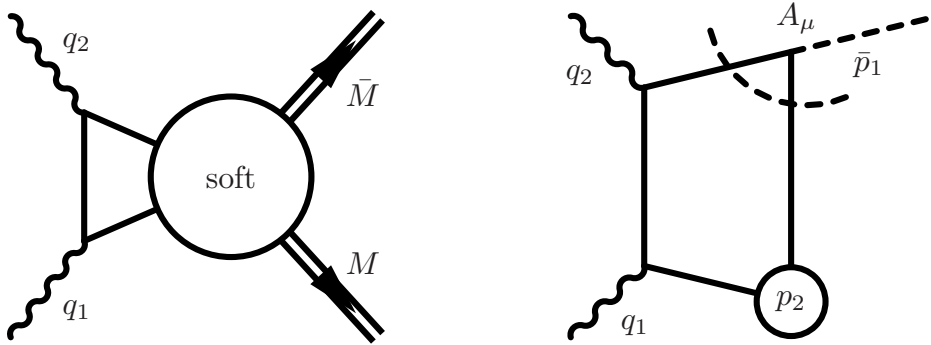}
\caption{a) the overall picture of the handbag contribution,\, b) the lowest
order  Feynman diagram for the light cone sum rule}
\end{figure}
Here, two photons interact with the same quark only, and these "active" ${\ov q}q$-quarks
carry nearly the whole meson momenta, while the additional "passive" ${\ov q^\prime}q^
\prime $ quarks are "wee partons" which are picked out from the vacuum by soft
non-perturbative interactions. It was obtained by DKV that the angular dependence of
amplitudes is $\sim 1/\sin^2\theta$ for all charged and neutral pions or kaons, while the
energy dependence is not predicted
and is described by some soft form factors $ R_{M}(s),$ which are then fitted to the data.
Because the "passive" quarks are picked out from the vacuum by soft non-perturbative forces,
these soft form factors are power suppressed at sufficiently large $s\,:  R_{M}(s) \leq 1/s^2
\,,\, $ in comparison with the leading meson form factors, $F_{M}(s)\sim 1/s\,$.

Some specific predictions of the handbag model look as\,:\\
 $\sigma_o(\pi^o\pi^o)/\sigma_o(\pi^-\pi^-)=1/2$,\, and
\bq
\frac{\sigma_o (K_S K_S)}{\sigma_o(K^+ K^-)}=\frac{1}{2}\Biggl |\frac{e^2_s\,R_{s\ra d}(s)
+e_d^2\,R_{d\ra s}(s)}{e_u^2\,R_{u\ra s}(s)+e_s^2\,R_{s\ra u}}  \Biggr |^2=\frac{2}{25}
\Biggl |\frac{1-\Delta}{1-\frac{8}{5}\Delta}\Biggr |^2 > 0.08.
\eq
Here, $e_u=2/3,\,\, e_d=e_s=-1/3$ are the quark charges, while the form factor $R_{u\ra s}
(s)$ corresponds to the active $u$-quark and passive $s$-quark, etc. It seems clear that it
is harder for soft interactions with the scale $\sim \Lambda_{QCD}$ to pick out from the
vacuum the heavier ${\ov s}s$-pair, than the light ${\ov u}u$-\, or ${\ov d}d$-pairs.
\footnote{\,\,\,
The effect due to $m_s\neq 0$ of the hard quark propagating between
two photons in fig.6 is small and can be neglected, see \cite{Ch}.
}
So\,:\,\, $ R_{u\ra s}(s)/R_{s\ra u}(s)\equiv (1-2\Delta)\,,\,\, \Delta > 0$.\,\,(\,The same
inequality $\Delta > 0$ follows from the fact that the heavier s-quark carries, on the
average, the larger fraction, $\langle x_s \rangle \,>\,0.5, $ of the K-meson momentum).
Therefore, the handbag model predicts that the number $2/25=0.080$ is the lower bound in
eq.(20).
\begin{figure}{\vspace*{-3.3 cm}}
\centering{\hspace*{-1.5 cm}}
\includegraphics[width=0.8\textwidth]{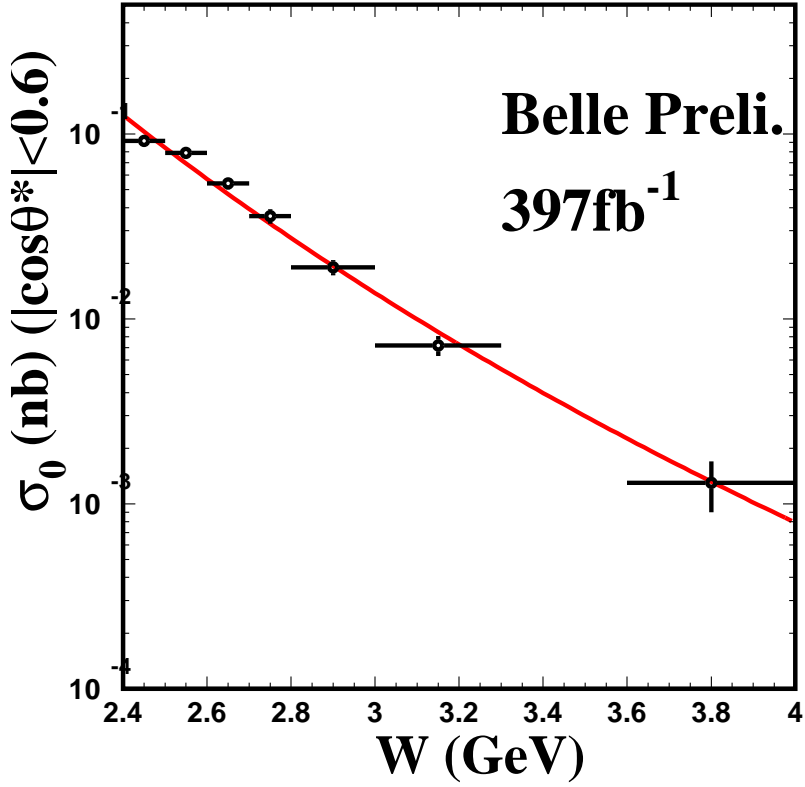}~
\caption{ The measured energy dependence of $\sigma_o(K_SK_S)$\,,\, \cite{Belle2}.
The solid line is $\,\sim W^{-k},\,\, k\simeq 10$.}
\end{figure}
\vspace{0.3cm}

The cross section $d\sigma (K_SK_S)/d\cos\theta $ has been measured recently by the Belle
collaboration \cite{Belle2}. The energy dependence at $2.4\,{\rm GeV} < W < 4.0\,{\rm GeV}$
was found to be\,:\, $\sigma_o(K_S K_S)\sim W^{-k}\,,\,k=(9.93\pm 0.4 )$, see fig.7\,. As for
the angular distribution, it is consisted with $d\sigma(K_S K_S)/d\cos\theta \sim 1/(W^{10}
\sin^4\theta)$, \cite{Belle2}. The measured ratio $(K_S K_S)/(K^+K^-)$ decreases strongly with
increasing energy, and becomes smaller than the lower bound $0.080$ in eq.(20) at $W\simeq 2.8
\,{\rm GeV}$. This is in contradiction with the hand-bag model
predictions.~\footnote{\,\, The ratio $\sigma_o(K_SK_S)/\sigma_o(K^+K^-)$ of measured cross
sections \cite{Belle1},\,\cite{Belle2} is\,: $\simeq 0.030$ at $W=3.2\,{\rm GeV}$,\,
and $\simeq 0.015$ at $W=3.8\,{\rm GeV}\,$, see fig.5b and fig.7.}\\

Moreover, the recent explicit calculation of the hand-bag diagrams in \cite{Ch} using the
method of the light cone sum rules \cite{Braun}\,\cite{NN}, see fig.6b\,,
shows that for all channels,
$\pi^+\pi^-\,,\, K^+K^-$ and $K_SK_S$\,: a) the energy dependence of the handbag form factors
$R_{M}(s)$ is $\sim (s_o/s)^2$ already in the energy interval $2.5\,{\rm GeV}< W=\sqrt s < 4\,
{\rm GeV}$ where the experiments have been done, and so this disagrees with the data on
$\sigma_o(\pi^+\pi^-)$ and $\sigma_O(K^+K^-)$ ;\, b) all handbag amplitudes do not depend on
the scattering angle $\theta$ (in contradiction with the DKV results)\, :
\bq
 \frac{d\sigma^{handbag}({\ov M}M)}{d\cos\theta}\sim
 \frac{|R_{M}(s)|^2}{s} \sim \frac{const}{W^{10}}\,,
\eq
and this disagrees with the data which show $(\sim 1/\sin^4\theta)$\, dependence\,;
c) and finally, the absolute values of all three cross sections predicted by the hanbag
model are much smaller than their experimental values.\\

The conclusion for this section is that the leading term QCD predictions for $\gamma\gamma
\ra {\ov M}M$ are in a reasonable agreement with the data (but only for the wide pion and
kaon wave functions, like $\phi_{\pi,K}^{(CZ)}(x)$), while the hand-bag model contradicts
the data in all respects\,: the energy dependence, the angular distribution and the
absolute normalization.

\end{document}